\documentclass[doublespace,times]{qjrms4}
\usepackage[colorlinks,bookmarksopen,bookmarksnumbered,citecolor=red,urlcolor=red]{hyperref}
\usepackage{moreverb}
\usepackage{empheq}
\usepackage{algorithm}
\usepackage{algpseudocode}

\usepackage{natbib}
\newcommand\BibTeX{{\rmfamily B\kern-.05em \textsc{i\kern-.025em b}\kern-.08em
T\kern-.1667em\lower.7ex\hbox{E}\kern-.125emX}}

\newcommand{\mb} {\mathbf}
\newcommand{\T}{^{\mathrm T}}

\def\volumeyear{2017}

\runningheads{P.~Sakov, J.-M.~Haussaire and M.~Bocquet}{An iterative EnKF in presence of additive model error}

\begin{document}

\title{An iterative ensemble Kalman filter \\ in presence of additive model error}
\author{Pavel Sakov\affil{a}\corrauth, Jean-Matthieu Haussaire\affil{b} and Marc Bocquet\affil{b}}
\corraddr{GPO Box 1538, Hobart, TAS 7000, Australia. E-mail: pavel.sakov@bom.gov.au}
\address{\affilnum{a}Environment and Research Division, Bureau of Meteorology, Melbourne, Australia\\
\affilnum{b}CEREA, Joint laboratory \'Ecole des Ponts ParisTech and EDF R\&D, Universit\'e Paris-Est, Champs-sur-Marne, France}

\begin{abstract}
  The iterative ensemble Kalman filter (IEnKF) in a deterministic framework was introduced in \citet{sak12a} to extend
  the ensemble Kalman filter (EnKF) and improve its performance in mildly up to strongly nonlinear cases.
  However, the IEnKF assumes that the model is perfect.  This assumption simplified the update of the system at a time
  different from the observation time, which made it natural to apply the IEnKF for smoothing.  In this study, we
  generalise the IEnKF to the case of imperfect model with additive model error.

  The new method called IEnKF-Q conducts a Gauss-Newton minimisation in ensemble space.  It combines the propagated
  analysed ensemble anomalies from the previous cycle and model noise ensemble anomalies into a single ensemble of
  anomalies, and by doing so takes an algebraic form similar to that of the IEnKF.  The performance of the IEnKF-Q is
  tested in a number of experiments with the Lorenz-96 model, which show that the method consistently outperforms both
  the EnKF and the IEnKF naively modified to accommodate additive model noise.
\end{abstract}


\maketitle

\section{Introduction}

  The analysis step in the Kalman filter \citep[KF,][]{kal60} can be seen as a single iteration of the Gauss-Newton
minimisation of a nonlinear cost function \citep{bel94a}.  It yields an exact solution in the linear case and works well
in weakly nonlinear cases, but becomes increasingly suboptimal as the system's nonlinearity increases.  The same
limitation applies to the ensemble Kalman filter \citep[EnKF,][]{eve94a}, which represents a state space formulation of
the KF suitable for large-scale applications.

To handle cases of stronger nonlinearity, a number of iterative EnKF schemes have been developed.  \citet{gu07a} introduced
the ensemble randomized maximum likelihood filter (EnRML) method, which represents a stochastic (Monte-Carlo)
Gauss-Newton solver.  \citet{sak12a} developed its deterministic analogue called the iterative EnKF (IEnKF) and tested
its performance in a number of significantly nonlinear situations with low-order models.  Both the EnRML and IEnKF do
essentially rely on the assumption that the model is perfect.  This assumption allows one to apply ensemble transforms
calculated in the course of data assimilation (DA) to the ensemble at the time of the previous analysis, as in the
ensemble Kalman smoother \citep[EnKS,][]{eve00a}, and then re-apply the forward model.  Transferring the ensemble
transforms back in time improves the initial estimates of the state and state error covariance, which in turn reduces
the nonlinearity of the system and improves the forecast (the background) and forecast covariance (the background
covariance) used in calculating the analysis at the next iteration.  Despite processing the same observations multiple
times, the IEnKF maintains the balance between the background and observation terms in the cost function: each next
iteration represents a correction to the previous one rather than a new assimilation of the same observations.  It is
different in this respect from the Running in Place scheme \citep[RIP,][]{kal10a,yan12a}, which adopts the latter
approach.  The RIP also has a stochastic implementation \citep{lor11a}.  A Bayesian derivation of the IEnKF, which
suggests its optimality for nonlinear chaotic models, has been given in section~2 of \citet{boc14a}.

The perfect model framework makes it possible to extend the IEnKF for assimilating future observations, or smoothing.
The corresponding method is known as the iterative ensemble Kalman smoother \citep[IEnKS,][]{boc14a,boc13a}.  The IEnKF
can also be enhanced to accommodate the inflation-less EnKF \citep[IEnKF-N,][]{boc12a}, and the ensemble space
formulation of the IEnKF algorithm makes it possible to localise it \citep{boc16a} with the localisation method known as
the local analysis \citep{eve03a}.  Moreover, it is possible to base the iterative EnKF on minimisation methods other
than the Gauss-Newton, e.g., on the Levenberg-Marquardt method \citep{boc12a,che13a}.

Along with the listed above single-cycle iterative schemes, there are also a variety of multi-cycle iterative EnKF
methods, emerging mostly from applications with static or quasi-static model state, such as oil reservoir modelling
\citep[e.g.,][]{li09a}.  Such methods involve re-propagation of the system from the initial time using the last
estimation of the static parameters of the model.  They are less suitable for applications with chaotic (e.g., with
atmospheric or oceanic) models, when the divergence at any single cycle can be typically considered as a crash of the
system.

The additive model error can be straightforwardly included into Monte-Carlo, or stochastic formulations of either
iterative or non-iterative EnKF schemes.  While it has not been formally considered in the original EnRML \citep{gu07a},
it was a part of the iterative ensemble smoother by \citet{man16a}.

Despite the intensive developments of the deterministic iterative EnKF schemes, so far they have not rigorously included
the model error.  One reason for that is the simplicity of the asynchronous DA in the perfect model framework
\citep{eve00a,hun04a,sak10a}.  The other reason is that the model error increases the dimension of the minimisation
problem: if in the perfect model case the optimal model state at any particular time defines the whole optimal model
trajectory, with a non-perfect model the global in time optimal solution represents a set of optimal model states at
each DA cycle.  This complicates the problem even in the simplest case of sequential DA considered in this study.

The development of an IEnKF framework with imperfect model can have a number of important theoretical and practical
implications.  Firstly, it can help understand limits of applicability of the perfect-model framework and limitations of
empirical treatments of the model error.  Further, it would be interesting to see whether/when adding empirically the
model error term to the cost function can have a regularising effect similar to that of the transition from the strong
constraint 4D-Var to the weak-constraint 4D-Var.  Including the model error has the potential to successfully address
situations when a large model error can be expected, such as of a probable algal bloom in biogeochemical models, or of a
rain event in land models.

This study develops an iterative method called IEnKF-Q based on the Gauss-Newton minimisation in the case of a system
with additive model error. In the following, the non-Gaussianity of the data assimilation system originates from the
nonlinearity of the model dynamics and the observation frequency. This data assimilation system is assumed to lie in a
range from a weakly nonlinear to a strongly nonlinear regime, where the EnKF might fail but where multimodality of the
underlying cost function is still not prominent. Nonetheless, by construction, the IEnKF-Q could also accommodate
nonlinear observation operators and to some extent, which is context-dependent, non-Gaussian variables due to its
variational analysis.  The strongly nonlinear regime where multimodality becomes prominent and where the iterative
ensemble Kalman filter and smoother could fail has been discussed in the conclusions of \citet{boc14a}. We refer to
\citet{fil17} for a more complete study of this strongly nonlinear regime but in a perfect model context.

The outline of the study is as follows.  The IEnKF-Q method is formulated in section~\ref{sec:problem}.
If the observation operator is linear, an alternative formulation resulting in the decoupling into a smoothing and a
filtering analysis is discussed in section~\ref{sec:decoupling}.  A pseudo-code for the IEnKF-Q algorithm is presented
in section~\ref{sec:algorithm}, and its performance with the Lorenz-96 model is tested in section~\ref{sec:tests}.
These tests include a preliminary study of a local IEnKF-Q. The results are discussed in section~\ref{sec:discussion}
and summarised in section~\ref{sec:summary}.

\section{Formulation and derivation}
\label{sec:problem}

The IEnKF-Q method is introduced from the more general context of the following global in time cost function $J_K$:
\begin{subequations}
  \label{global}
  \begin{align}
    \{\mb x_i^\star\}_{i=1}^K = & \arg \underset{\{\mb x_i\}_{i = 1}^K}\min J_K(\mb x_1, \dots, \mb x_K), \\
    \nonumber
     J_K(\mb x_1, \dots, \mb x_K) =&  \frac{1}{2}\|\mb x_1 - \mb x_1^\mathrm{f}\|^2_{(\mb P_1^\mathrm{f})^{-1}} + \frac{1}{2}\sum_{i = 2}^K \|\mb y_i  - \mathcal H_i(\mb x_i)\|^2_{\mb R_i^{-1}} \\
     & + \frac{1}{2}\sum_{i = 2}^K\|\mb x_i - \mathcal M_i(\mb x_{i-1})\|^2_{\mb Q_i^{-1}}. 
  \end{align}
\end{subequations}
Here $i$ is the cycle number associated with time, $K$ -- number of cycles plus one, $\mb x_i$ -- (model) state at cycle $i$,
$\mb x_i^\star$ -- state estimate (analysis) at cycle $i$, $\mb x_1^\mathrm{f}$ -- initial state estimate, $\mb P_1^\mathrm{f}$ -- initial
state error covariance, $\mb y_i$ -- observations, $\mathcal H_i$ -- observation operator, $\mb R_i$ -- observation
error covariance, $\mathcal M_i$ -- model operator, and $\mb Q_i$ - model error covariance; and the norm notation $\|
\mb x \|^2_{\mb B} \equiv \mb x\T \mb B \mb x$ is used.  The cost function is assumed to be, generally, nonlinear due to
nonlinear operators $\mathcal M$ and $\mathcal H$.

In the case of linear $\mathcal M$ and $\mathcal H$ the problem (\ref{global}) becomes quadratic and has recursive
solutions.  The last component of the solution $\mb x_K^\star$ is known as the filtering analysis and is given by the KF,
while the whole analysis $\{\mb x_i^\star\}_{i=1}^K$ is given by the Kalman smoother.  In the nonlinear case, it is
essential for the applicability of recursive, or sequential, methods based on the KF, such as the EnKF, that the
nonlinearity of the system is weak.  The rationale for iterative methods such as the IEnKF is that the weak nonlinearity
needs to be achieved only in the course of minimisation; then the final analysis is calculated essentially for a linear
system.

Therefore, the focus of an iterative method is a \emph{single} analysis cycle (i.e. $K=2$) and the associated problem that arises in the
course of the iterative solution of
\begin{subequations}
  \label{local}
  \begin{align}
    \{\mb x_1^\star, \mb x_2^\star\} =& \arg \underset{\{\mb x_1, \mb x_2\}}\min J(\mb x_1, \mb x_2), \\
    \nonumber
    J(\mb x_1, \mb x_2) =& \frac{1}{2}\|\mb x_1 - \mb x_1^\mathrm{a}\|^2_{(\mb P_1^\mathrm{a})^{-1}} + \frac{1}{2}\|\mb y_2 - \mathcal H(\mb x_2)\|^2_{\mb R^{-1}}\\
    \label{J-local}
    & + \frac{1}{2}\|\mb x_2 - \mathcal M(\mb x_1)\|^2_{\mb Q^{-1}}.
  \end{align}
\end{subequations}
Here we have dropped the absolute time indices and use relative indices 1 and 2, which refer to analysis times $t_1$ and
$t_2$; $\mb x_1^\mathrm{a}$ and $\mb P_1^\mathrm{a}$ are the filtering analysis and filtering state error covariance at
time $t_1$, which have been obtained in the previous cycle; all other variables have direct analogues in formulation
(\ref{global}) of the global problem. The function \eqref{J-local} should be seen as the state-space cost function
associated with the analysis of the IEnKF-Q. It is the key to the method's derivation.

The main difference between the KF and the EnKF is their representation of the state of the DA system (SDAS).  In the
KF, the SDAS is carried by the state estimate $\mb x$ and state error covariance $\mb P$.  In the EnKF, the SDAS is
carried by an ensemble of model states $\mb E$.  These two representations are related as follows:
\begin{subequations}
  \label{enkf}
  \begin{align}
    \label{x-enkf}
    & \mb x = \mb E \mb 1 / m,\\
    \label{A-enkf}
    & \mb P = \mb A \mb A\T / (m-1),\\
    & \mb A \equiv \mb E - \mb x \mb 1\T = \mb E \, (\mb I - \mb 1 \mb 1\T / m),
  \end{align}
\end{subequations}
where $m$ is the ensemble size, and $\mb 1$ is a vector with all components equal to 1.

The problem formalised by (\ref{local}) can be solved by finding zero gradient of the cost function (\ref{J-local}):
\begin{align}
  \left\{
  \begin{array}{l}
    \nabla_{\mb x_1} J(\mb x_1^\star, \mb x_2^\star) = \mb 0,\\
    \nabla_{\mb x_2} J(\mb x_1^\star, \mb x_2^\star) = \mb 0,\\
  \end{array}
  \right .
\end{align}
similarly to the approach in \citet{sak12a}.
However, for an ensemble-based system the derivation becomes simpler if the solution is sought in ensemble space.
Let us assume
\begin{subequations}
  \label{u}
  \begin{align}
    \label{x1}
    & \mb x_1 = \mb x_1^\mathrm{a} + \mb A_1^\mathrm{a} \mb u, \\
    \label{A-ienkfq}
    & \mb A_1^\mathrm{a} (\mb A_1^\mathrm{a})\T = \mb P_1^\mathrm{a},\\
    & \mb A_1^\mathrm{a} \mb 1 = \mb 0,
\end{align}
\end{subequations}
where $\mb A_1^\mathrm{a}$ is defined as the matrix of the centred anomalies resulting from a previous analysis at $t_1$, and
\begin{subequations}
  \label{v}
  \begin{align}
    \label{x2}
    & \mb x_2 = \mathcal M(\mb x_1) + \mb A_2^q \mb v,\\
    \label{A2q}
    & \mb A_2^q (\mb A_2^q)\T = \mb Q,\\
    & \mb A_2^q \mb 1 = \mb 0.
  \end{align}
\end{subequations}
We seek solution in $(\mb u, \mb v)$ rather than in $(\mb x_1, \mb x_2)$ space.
Note that for convenience we use a different normalisation of the ensemble anomalies in (\ref{A-ienkfq}) than in (\ref{A-enkf}).
After substituting (\ref{u},\ref{v}) into (\ref{local}) the problem takes the form
\begin{subequations}
  \begin{align}
    &\{\mb u^\star,\mb v^\star\} = \arg \min_{\{\mb u, \mb v\}} J(\mb u, \mb v),\\
    &J(\mb u, \mb v) = \frac{1}{2}\mb u\T \mb u + \frac{1}{2}\left\| \mb y_2 - \mathcal H(\mb x_2) \right\|^2_{\mb R^{-1}} + \frac{1}{2}\mb v\T \mb v,
  \end{align}
\end{subequations}
or, concatenating $\mb u$ and $\mb v$,
\begin{subequations}
  \begin{align}
    \label{w}
    & \mb w \equiv \left[ \begin{array}{c} \mb u \\ \mb v \end{array} \right], \\
    & \mb w^\star = \arg \min_{\mb w} J(\mb w),\\
    \label{J}
    & J(\mb w) = \frac{1}{2}\mb w\T \mb w + \frac{1}{2}\left\| \mb y_2 - \mathcal H(\mb x_2) \right\|^2_{\mb R^{-1}},
  \end{align}
\end{subequations}
where according to (\ref{u}), (\ref{v}) and (\ref{w})
\begin{align}
  \mb x_2(\mb w) = \mathcal M(\mb x_1^\mathrm{a} + \mb A_1^\mathrm{a} \mb w_{1:m}) + \mb A_2^q \mb w_{m+1:m+m_q},
\end{align}
$m_q$ is the size of the model noise ensemble $\mb A_2^q$, and $\mb w_{n_1:n_2}$ denotes a subvector of $\mb w$ formed
by elements from $n_1$ to $n_2$.  Condition of zero gradient of the cost function (\ref{J}) yields
\begin{align}
  \label{zerograd}
    \mb w - (\mb H \mb A)\T \mb R^{-1} \left[ \mb y_2 - \mathcal H(\mb x_2) \right] = \mb 0,
\end{align}
where
\begin{align}
  \label{A}
  & \mb A \equiv [\mb M \mb A_1^\mathrm{a}, \mb A_2^q],\\
  \label{H}
  & \mb H \equiv \nabla \mathcal H(\mb x_2),\\
  \label{M}
  & \mb M \equiv \nabla \mathcal M(\mb x_1).
\end{align}

Equation (\ref{zerograd}) can be solved iteratively by the Newton method:
\begin{align}
  \label{newton}
  \mb w^{i+1} = \mb w^i - \mb D^i \nabla J(\mb w^i),
\end{align}
where $\mb D^i$ is the inverse Hessian of the cost function (\ref{J}), and hereafter index $i$ denotes the value of the
corresponding variable at the $i$th iteration.  We ignore the second-order derivatives in calculating the Hessian, which
corresponds to employing the Gauss-Newton minimisation, so that
\begin{align}
  \label{D}
  \mb D^i \approx \left[ \mb I + (\mb H^i \! \mb A^{\! i})\T \mb R^{-1} \mb H^i \! \mb A^{\! i} \right]^{-1},
\end{align}
and \eqref{newton} becomes
\begin{align}
  \label{w-sol}
   \mb w^{i+1} - \mb w^i = & \left[ \mb I + (\mb H^i \! \mb A^{\! i})\T \mb R^{-1} \mb H^i \! \mb A^{\! i} \right]^{-1}  \nonumber \\
  &  \times \left\{(\mb H^i \! \mb A^{\! i})\T \mb R^{-1} \left[\mb y_2 - \mathcal H(\mb x_2^i)\right] - \mb w^i \right\}.
\end{align}
This equation, required to obtain the analysis state, is the core of the IEnKF-Q method.

The other two necessary elements of the IEnKF-Q are the computations of the smoothed and filtered ensemble anomalies $\mb
A_1^\mathrm{s}$ and $\mb A_2^\mathrm{a}$.  Knowledge of $\mb A_1^\mathrm{s}$ is needed to reduce the ensemble spread at
$t_1$ in accordance with the reduced uncertainty after assimilating observations at $t_2$; and $\mb A_2^\mathrm{a}$ is
needed to commence the next cycle.

To find the analysed ensemble anomalies at $t_2$, we first define the perturbed states at $t_1$ and $t_2$:
\begin{equation}
\label{eq:perturbations}
  \delta \mb x_1 = \mb A_1^\mathrm{a} \, \delta \mb u, \qquad  \delta \mb x_2 = \mb M \mb A_1^\mathrm{a} \, \delta \mb u + \mb A_2^q \, \delta \mb v ,
\end{equation}
in terms of the perturbed $\delta \mb u$ and $\delta \mb v$.  We note that in the linear case, $\mb
D^\star$ approximated by (\ref{D}) represents the covariance in ensemble space:
\begin{align}
  \label{P_w}
  \mathrm{E}[\mb w^\star (\mb w^\star)\T] = \mb D^\star,
\end{align}
where $\mathrm{E}$ is the statistical expectation and index $\star$ denotes the value of the corresponding variable after
convergence, so that, using \eqref{eq:perturbations}:
\begin{align}
  \mb A_2^\mathrm{a} (\mb A_2^\mathrm{a})\T &=\mathrm{E}[\delta \mb x_2^\star (\delta \mb x_2^\star)\T] \nonumber \\
  &= \mb A^{\! \star} \mathrm{E}[\mb w^\star (\mb w^\star)\T] (\mb A^{\! \star})\T = \mb A^{\! \star} \mb D^\star (\mb A^{\! \star})\T,
\end{align}
and
\begin{align}
  \nonumber
  \mb A_2^\mathrm{a} &= \mb A^{\! \star} (\mb D^\star)^{1/2}\\
  \label{A2a}
  & = \mb A^{\! \star} \! \left[ \mb I + (\mb H^\star \! \mb A^{\! \star})\T (\mb R)^{-1} \mb H^\star \! \mb A^{\! \star} \right]^{-1/2},
\end{align}
where $\mb D^{1/2}$ is the unique symmetric positive (semi-)definite square root of a positive (semi-)definite matrix $\mb D$.

Similarly,
\begin{equation}
  \mb A_1^\mathrm{s} (\mb A_1^\mathrm{s})\T = \mathrm{E}[\delta \mb x_1^\star (\delta \mb x_1^\star)\T]
  = \mb A_1^\mathrm{a} \mathrm{E}[\mb u^\star (\mb u^\star)\T] (\mb A_1^\mathrm{a})\T;
\end{equation}
therefore
\begin{align}
  \label{A1s}
  \mb A_1^\mathrm{s} = \mb A_1^\mathrm{a} \, (\mb D^\star_{1:m,1:m})^{1/2},
\end{align}
where $\mb D_{n_1:n_2,m_1:m_2}$ denotes a submatrix of $\mb D$ formed by rows from $n_1$ to $n_2$ and columns from $m_1$
to $m_2$.  It can be verified using (\ref{Du}) that the smoothed error covariance $\mb P_1^\mathrm{s} = \mb
A_1^\mathrm{s} (\mb A_1^\mathrm{s})\T$ matches the Kalman smoother solution \citep[][eq.~3.31]{rau65a}.

Equations (\ref{w-sol}), (\ref{A2a}) and (\ref{A1s}) constitute the backbone of the IEnKF-Q.

\section{Decoupling of $\mb u$ and $\mb v$ in the case of linear observations}
\label{sec:decoupling}

In the case of a linear observation operator $\mathcal H$, it is possible to decouple the solution for $\mb u$ and $\mb v$.
This is shown below by transforming (\ref{w-sol}) to an alternative form.

Re-writing (\ref{w-sol}) as
\begin{align}
  \mb w^{i+1}= & \left[ \mb I + (\mb H^i \! \mb A^{\! i})\T \mb R^{-1} \mb H^i \! \mb A^{\! i} \right]^{-1}   \nonumber \\
  & \times (\mb H^i \! \mb A^{\! i})\T \mb R^{-1} \left[\mb y_2 - \mathcal H(\mb x_2^i) + \mb H^i \! \mb A^{\! i} \mb w^i \right]
\end{align}
and using the identity
\begin{align}
  \left[ \mb I + \mb B\T \mb R^{-1} \mb B \right]^{-1} \mb B\T \mb R^{-1} = \mb B\T (\mb B \mb B\T + \mb R)^{-1},
\end{align}
where $\mb R$ is positive definite, we get
\begin{align}
   \mb w^{i+1} =& (\mb H^i \! \mb A^{\! i})\T \left[(\mb H^i \! \mb A^{\! i})\T \mb H^i \! \mb A^{\! i} + \mb R\right]^{-1} \nonumber \\
   & \times \left[\mb y_2 - \mathcal H(\mb x_2^i) + \mb H^i \! \mb A^{\! i} \mb w^i\right],
\end{align}
or, decomposing $\mb w^i$ and $\mb A^i$,
\begin{empheq}[left=\empheqlbrace]{alignat=1}
    \mb u^{i+1} =&  (\mb H^i \mb M^i \! \mb A_1^\mathrm{a})\T \left[(\mb H^i \mb M^i \! \mb A_1^\mathrm{a})\T \mb H^i \mb M^i \! \mb A_1^\mathrm{a} + \mb R_u^i\right]^{-1} \nonumber \\
    & \times \left[\mb y_2 - \mathcal H\left(\mb x_2^i\right) + \mb H^i \mb M^i \! \mb A_1^\mathrm{a} \mb u^i + \mb H^i \! \mb A_2^q \mb v^i\right],\\
    \mb v^{i+1} =& (\mb H^i \! \mb A_2^q)\T \left[(\mb H^i \! \mb A_2^q)\T \mb H^i \! \mb A_2^q + \mb R_v^i\right]^{-1} \nonumber \\
    & \times \left[\mb y_2 - \mathcal H\left(\mb x_2^i\right) + \mb H^i \mb M^i \! \mb A_1^\mathrm{a} \mb u^i + \mb H^i \! \mb A_2^q \mb v^i\right],
\end{empheq}
where
\begin{align}
  \label{Ru}
    & \mb R_u^i \equiv \mb H^i \! \mb A_2^q (\mb H^i \! \mb A_2^q)\T + \mb R,\\
  \label{Rv}
  & \mb R_v^i \equiv \mb H^i \mb M^i \! \mb A_1^\mathrm{a} (\mb H^i \mb M^i \! \mb A_1^\mathrm{a})\T + \mb R.
\end{align}
Focusing on the increments of the iterates, we equivalently obtain
\begin{empheq}[left=\empheqlbrace]{alignat=1}
  \nonumber
  \mb u^{i + 1} - \mb u^i =& \mb D_u^i \left\{(\mb H^i \mb M^i \mb A_1^\mathrm{a})\T (\mb R_u^i)^{-1} \right.\\
  \label{u-sol}
  & \left. \times \left[\mb y_2 - \mathcal H(\mb x_2^i) + \mb H^i \! \mb A_2^q \mb v^i\right] - \mb u^i\right\},\\
  \nonumber
  \mb v^{i + 1} - \mb v^i =& \mb D_v^i \left\{(\mb H^i\! \mb A_2^q)\T (\mb R_v^i)^{-1} \right.\\
  \label{v-sol}
   & \left. \times \left[\mb y_2 - \mathcal H(\mb x_2^i) + \mb H^i \mb M^i \! \mb A_1^\mathrm{a} \mb u^i\right] - \mb v^i\right\},
\end{empheq}
where
\begin{align}
  \label{Du}
  &\mb D_u^i \equiv \left[\mb I + (\mb H^i \mb M^i \! \mb A_1^\mathrm{a})\T (\mb R_u^i)^{-1} \mb H^i \mb M^i \! \mb A_1^\mathrm{a}\right]^{-1},\\
  \label{Dv}
  &\mb D_v^i \equiv \left[\mb I + (\mb H^i \! \mb A_2^q)\T (\mb R_v^i)^{-1} \mb H^i \! \mb A_2^q\right]^{-1}.
\end{align}
It is straightforward to verify that $\mb D_u^i = \mb D^i_{1:m,1:m}$, and $\mb D_v^i = \mb D^i_{m+1:m+m_q,m+1:m+m_q}$.

Equations (\ref{u-sol}) and (\ref{v-sol}) represent an alternative form of equation (\ref{w-sol}) that makes it easy to
see the decoupling of $\mb u$ and $\mb v$ in the case of linear $\mathcal H$.  In this case $\mb H^i = \mb H =
\mathrm{Const}$ and
\begin{equation}
  \mathcal H(\mb x_2^i) = \mathcal H [\mathcal M(\mb x_1^i) + \mb A_2^q \mb v^i ] =  \mathcal H \circ \mathcal M(\mb x_1^i) + \mb H \! \mb A_2^q \mb v^i,
\end{equation}
so that (\ref{u-sol}) becomes
\begin{align}
  \label{u-lin}
  \mb u^{i + 1} - \mb u^i = & \mb D_u^i \left\{(\mb H \mb M^i \! \mb A_1^\mathrm{a})\T (\mb R_u^i)^{-1} \right. \nonumber \\
  &\left. \times \left[\mb y_2 - \mathcal H \circ \mathcal M(\mb x_1^\mathrm{a} + \mb A_1^\mathrm{a} \mb u^i)\right] - \mb u^i\right\}.
\end{align}
It follows from \eqref{u-lin} that in the case of linear observations, $\mb u$ can be found by the IEnKF algorithm (i.e,
assuming perfect model) with modified observation error (\ref{Ru}): $\mb R^i_u = \mb R + \mb H \mb Q \mb H\T$.  After that, $\mb v$ can be found from \eqref{v-sol}:
\begin{equation}
  \label{v-lin}
  \mb v^\star = (\mb H\! \mb A_2^q)\T (\mb R_v^\star)^{-1} \left[\mb y_2 - \mathcal H(\mb x_2^\star) + \mb H \mb M^\star \! \mb A_1^\mathrm{a} \mb u^\star\right],
\end{equation}
which can further be simplified using $\mb x_2^\star = \mathcal M(\mb x_1^\star) + \mb A_2^q \mb v^\star$  and
\begin{equation}
  \label{eq:u-dec}
  \mb u^\star = (\mb H \mb M^\star \! \mb A_1^\mathrm{a})\T (\mb R_u^\star)^{-1}\left[\mb y_2 - \mathcal H \circ \mathcal M(\mb x_1^\star) \right]
\end{equation}
obtained from  \eqref{u-lin}, finally yielding the non-iterative estimator
\begin{equation}
  \label{eq:v-dec}
  \mb v^\star = (\mb H\! \mb A_2^q)\T(\mb R + \mb H \mb Q \mb H\T)^{-1}\left[\mb y_2 - \mathcal H \circ \mathcal M(\mb x_1^\star) \right].
\end{equation}

Computationally, the decoupling reduces the size of $\mb D^i$, i.e. $(m + m_q) \times (m + m_q)$, to that of $\mb
D_u^i$, i.e. $m \times m$; however, it involves the inversion of a $p \times p$ matrix $\mb R_u^i$, where $p$ is the
number of observations, which in large-scale geophysical systems can be expected to be much larger than the ensemble
sizes $m$ and $m_q$.

The decoupling of $\mb u$ and $\mb v$ can be analysed in terms of probability distributions.  This allows one to
understand it at a more fundamental level and to connect the IEnKF-Q to the particle filter with optimal proposal
importance sampling \citep{dou00a}, which is an elegant particle filter solution of our original problem with
applications to the data assimilation in geosciences \citep{boc10a,sny15a,sli16a}.

The posterior probability density function (PDF) of the analysis $p(\mb x_1, \mb x_2 | \mb y_2)$ is related to the
IEnKF-Q cost function \eqref{J-local} through
\begin{align}
  J(\mb x_1,\mb x_2) = -\ln p(\mb x_1, \mb x_2 | \mb y_2).
\end{align}
In all generality, the posterior PDF can be decomposed into
\begin{align}
  p(\mb x_1, \mb x_2 | \mb y_2) = p(\mb x_2 | \mb x_1 , \mb y_2) p(\mb x_1 | \mb y_2).
\end{align}
It turns out that when $\mathcal H$ is linear both factors of this product have an analytic expression.
This simplification is leveraged over when defining a particle filter with an optimal importance sampling \citep{dou00a}.
For our problem, one can show after some elementary but tedious matrix algebra that
\begin{align}
 &-\ln p(\mb x_1 | \mb y_2) = \frac{1}{2}\left\|\mb x_1 - \mb x_1^\mathrm{a} \right\|^2_{(\mb P_1^\mathrm{a})^{-1}} \nonumber \\
 & \qquad + \frac{1}{2}\left\|\mb y_2 - \mathcal H \circ \mathcal M(\mb x_1) \right\|^2_{(\mb R + \mb H \mb Q \mb H\T)^{-1}} + c_1 ,
\end{align}
and
\begin{align}
  \nonumber
  & -\ln p(\mb x_2 | \mb x_1 , \mb y_2) = \\
  \nonumber
  &\qquad \frac{1}{2}\|\mb x_2 - \mathcal M(\mb x_1) - \mb Q \mb H\T (\mb R + \mb H \mb Q \mb H\T )^{-1}\\
  &\qquad \times \left[\mb y_2 - \mathcal H  \circ \mathcal M(\mb x_1)\right] \|^2_{\mb Q^{-1} + \mb H\T \mb R^{-1} \mb H} + c_2 ,
\end{align}
where $c_1$ and $c_2$ are constants that neither depend on $\mb x_1$ nor $\mb x_2$.  Note that $p(\mb x_1 | \mb y_2)$ is
non-Gaussian while $p(\mb x_2 | \mb x_1 , \mb y_2)$ is a Gaussian PDF thanks to the linearity of $\mb H$.

This decomposition enables to minimise $J(\mb x_1,\mb x_2)$ in two steps.  First, one can minimise $-\ln p(\mb x_1 |
\mb y_2)$ over $\mb x_1$ yielding the maximum a posteriori (MAP) solution $\mb x_1^\star$.  This identifies with the
smoothing analysis of a perfect model IEnKF but with $\mb R$ replaced with $\mb R + \mb H \mb Q \mb H\T$.  Second, the
MAP solution of the minimisation of $-\ln p(\mb x_2 | \mb x^\star_1 , \mb y_2)$ is directly given by
\begin{align}
\label{eq:MAP2}
  \mb x_2^\star =& \mathcal M(\mb x_1^\star) + \mb Q \mb H\T\left(\mb R + \mb H \mb Q \mb H\T \right)^{-1} \nonumber \\
  & \times \left[\mb y_2 - \mathcal H \circ \mathcal M(\mb x_1^\star)\right] .
\end{align}
It is simple to check that this expression, albeit written in ensemble space, is consistent with \eqref{eq:v-dec}.

This decomposition explains at a fundamental level why the computation of the MAP of the IEnKF-Q were to decouple in
(\ref{u-lin},\ref{eq:v-dec}) when the observation operator $\mathcal H$ is linear.  However, this decoupling, valid for
the MAP, does not immediately convey to the computation of the posterior perturbations.

\section{The base algorithm}
\label{sec:algorithm}

In this section, we put up an IEnKF-Q algorithm based on equations (\ref{w-sol}), (\ref{A2a}) and (\ref{A1s}).  We refer
to it as the base algorithm, because there are many possible variations of the algorithm based on different
representations of these equations, including using decoupling of $\mb u$ and $\mb v$ in the case of linear observations
described in section~\ref{sec:decoupling}.  Further, we do not include localisation, which is a necessary attribute of
large-scale systems.  The localisation of the IEnKF and IEnKS has been explored in \citet{boc16a}.  In this paper, an
implementation based on the local analysis method \citep{eve03a, sak11a} has been proposed and may require the use of a
surrogate model, typically advection by the fluid, to propagate a dynamically covariant localisation over long data
assimilation windows.  Such an implementation is actually rather straightforward for the IEnKF-Q since it is already
formulated in ensemble space. Even though this is not the focus of this study, we will make preliminary tests of a
local variant of the IEnKF-Q at the end of section~\ref{sec:tests} and provide its algorithm in Appendix A.

While the EnKF is a derivative-less method, it is possible to vary the type of approximations of Jacobians $\mb M$ and
$\mb H$ with the ensemble used in the algorithm.  In various types of the EnKF, it is common to use approximations of
various products of $\mb H$ and $\mb M$ using ensemble of finite spread set based on statistical estimation
(\ref{A-enkf}) for sample covariance:
\begin{subequations}
  \label{appr-enkf}
  \begin{align}
    &\mb H \mb x &\leftarrow & \quad \mathcal H (\mb E) \, \mb 1 / m, \\
    \label{HA}
    &\mb H \mb A &\leftarrow & \quad \mathcal H(\mb E)(\mb I - \mb 1 \mb 1\T / m),\\
    &\mb M \mb x &\leftarrow & \quad \mathcal M (\mb E) \, \mb 1 / m,\\
    &\mb M \mb A &\leftarrow & \quad \mathcal M(\mb E)(\mb I - \mb 1 \mb 1\T / m),\\
    &\mb H \mb M \mb A &\leftarrow & \quad \mathcal H \circ \mathcal M(\mb E) (\mb I - \mb 1 \mb 1\T /m ).
  \end{align}
\end{subequations}
However, as pointed in \citet{sak12a}, it is also possible to use finite difference approximations:
\begin{subequations}
  \label{appr-ekf}
  \begin{align}
    &\mb H \mb x &\leftarrow& \quad \mathcal H (\mb x \mb 1\T + \varepsilon \mb A) \, \mb 1/m, \\
    &\mb H \mb A &\leftarrow& \quad \mathcal H(\mb x \mb 1\T + \varepsilon \mb A)(\mb I - \mb 1 \mb 1\T / m) / \varepsilon,\\
    &\mb M \mb x &\leftarrow& \quad \mathcal M (\mb x \mb 1\T + \varepsilon \mb A) \, \mb 1/m,\\
    &\mb M \mb A &\leftarrow& \quad \mathcal M(\mb x \mb 1\T + \varepsilon \mb A)(\mb I - \mb 1 \mb 1\T / m) / \varepsilon,\\
    &\mb H \mb M \mb A &\leftarrow& \quad \mathcal H \circ \mathcal M(\mb x \mb 1\T + \varepsilon \mb A)(\mb I - \mb 1 \mb 1\T / m) / \varepsilon,
  \end{align}
\end{subequations}
where $\varepsilon \ll 1$.  Using these approximations results in methods of derivative-less state-space extended Kalman
filter (EKF) type.  The difference in employing approximations (\ref{appr-enkf}) and (\ref{appr-ekf}) is somewhat
similar to the difference between secant and Newton methods.  It is also possible to mix these two approaches by
choosing an intermediate value of parameter $\varepsilon$ in (\ref{appr-ekf}), e.g., $\varepsilon = 0.5$.

Approximations of EnKF and EKF types (\ref{appr-enkf}) and (\ref{appr-ekf}) were compared in a number of numerical
experiments in \citet{sak12a}.  It was found that generally using finite spread approximations (\ref{appr-enkf}) results
in more robust and better performing schemes.

It was found later \citep{boc12a} that performance of schemes based on finite difference approximations can be improved
by conducting a final propagation with a finite spread ensemble.  The corresponding schemes were referred to as
``bundle'' variants, while the schemes using finite spread approximations -- as ``transform'' variants.  The
algorithm~\ref{alg:ienkfq} is a transform variant of the IEnKF-Q method.

\begin{algorithm}
  \setstretch{1.35}
  \caption{\label{alg:ienkfq} A ``transform'' variant of the IEnKF-Q. The pieces of pseudo-code highlighted in red show
    changes relative to the IEnKF algorithm in absence of model error. ``$\mathrm{SR}(\mb A, m)$'' denotes ensemble size
    reduction from $m+m_q$ to $m$.}
  \begin{algorithmic}[1]
    \Function{$[\mb E_2]$ = \mbox{\bf ienkf\_cycle}}{$\mb E^\mathrm{a}_1{\color{red}, \, \mb A_2^q}, \, \mb y_2,\ \mb R, \, \mathcal M, \mathcal H$}
    \State $\mb x_1^\mathrm{a} = \mb E_1^\mathrm{a} \, \mb 1 / m$
    \State $\mb A_1^\mathrm{a} = (\mb E_1^\mathrm{a} - \, \mb x_1^\mathrm{a} \mb 1\T) / \sqrt{m - 1}$
    \State $\mb D = \mb I, \quad \mb w = \mb 0$
    \Repeat
    \State $\mb x_1 = \mb x_1^\mathrm{a} + \mb A_1^\mathrm{a} \mb w_{\color{red} 1 : m}$
    \State $\mb T = (\mb D_{\color{red} 1:m,1:m})^{1/2}$
    \State $\mb E_1 = \mb x_1 \mb 1\T + \mb A_1^\mathrm{a} \mb T \sqrt{m - 1}$
    \State $\mb E_2 = \mathcal M (\mb E_1)$
    \State $\mb {HA}_2 = \mathcal H(\mb E_2)(\mb I - \mb 1\mb 1\T/ \, m) \, \mb T^{-1}/ \sqrt{m - 1}$
    \State ${\color{red} \mb H \mb A_2^q = \mathcal H(\mb E_2 \mb 1 \mb 1\T/ \, m + \mb A_2^q \sqrt{m_q - 1})}$
    \Statex \hspace{3.5cm} ${\color{red} \times (\mb I - \mb 1 \mb 1\T / \, m_q) / \sqrt{m_q - 1}}$
    \State $\mb H \mb A = [\mb H \mb A_2{\color{red} , \mb H \mb A_2^q}]$
    \State $\mb x_2 = \mb E_2 \mb 1/m {\color{red} + \mb A_2^q \mb w_{m+1:m+m_q}}$
    \State $\nabla \! J =  \mb w - (\mb {HA})\T \mb R^{-1} [\mb y_2 - \mathcal H(\mb x_2)]$
    \State $\mb D = [\mb I + (\mb{HA})\T \mb R^{-1} \mb {HA}]^{-1}$
    \State $\Delta \mb w = -\mb D \, \nabla \! J$
    \State $\mb w := \mb w + \Delta \mb w$
    \Until $\| \Delta \mb w \| < \varepsilon$
    \State $\mb A_2 = \mb E_2 \, (\mb I - \mb 1 \mb 1\T / \, m) {\color{red} \, \mb T^{-1}}$
    \State ${\color{red}\mb A = [\mb A_2 / \sqrt{m - 1}, \mb A_2^q] \, \mb D^{1/2}}$
    \State ${\color{red}\mb A_2 = \mathrm{SR}(\mb A, m) \sqrt{m - 1}}$
    \State $\mb E_2 = \mb x_2 \mb 1\T + \mb A_2$
    \EndFunction
  \end{algorithmic}
  \end{algorithm}
Line 6 of the algorithm corresponds to (\ref{x1}); line 7 calculates the ensemble transform in (\ref{A1s});
multiplication by $\sqrt{m-1}$ on line 8 restores normalisation of ensemble anomalies before propagation to
statistically correct magnitude; division by $\sqrt{m-1}$ on line 10 changes it to the algebraically convenient form
$\mb P = \mb A \mb A\T$ used in (\ref{u}).  The observation ensemble anomalies of the model noise ensemble $\mb H^i \!
\mb A_2^q$ are calculated on line 11.  This involves adding ensemble mean and re-normalisation before applying the
observation operator.  In the case of linear observations this line would reduce to $\mb H \mb A_2^q = \mathcal H (\mb
A_2^q)$.  Line 13 corresponds to (\ref{x2}), and line 20 to (\ref{A2a}).

The ensemble transform applied in line 7 is actually a bit restrictive, though it is sleek and convenient.  Its
potential suboptimality is obvious in that the transform correctly applies to the evolution model propagation of the
ensemble, but not to the observation operator.  In this context, faithfully enforcing the transform principle proposed
in \citet{sak12a} would imply applying (on the right) the transform matrix $\mb T = \mb D^{1/2}$ to the joint anomaly
matrix $\left[ \mb A_1^\mathrm{a}, \, \mb A_2^q \right]$, before applying the nonlinear map from the ensemble space to
the observation space:
\begin{equation}
\mb w \mapsto \mathcal H\left( \mathcal M(\mb x_1^\mathrm{a} + \mb A_1^\mathrm{a} \mb w_{1:m}) + \mb A_2^q \mb
w_{m+1:m+m_q} \right).
\end{equation}
The implementation of this joint transform is less simple than that offered by the one in line 7, which merely amounts
to using the smoothing anomalies marginalised at $t_1$.  Another simple possibility is to choose $\mb T = [\mb
D^{1/2}]_{ 1:m,1:m}$, which remains a positive definite matrix.  We have checked that these three approaches yield the
same quantitative results for all the experiments reported below, except for those on localisation.  However, we expect
that the optimal joint transform mentioned above could make a difference in the presence of a significantly nonlinear
observation operator (not tested).

Because the IEnKF-Q uses augmented ensemble anomalies (\ref{A})\footnote{The augmentation of the propagated state error
  anomalies and model error anomalies has also been used in the reduced rank square root filter by
  \citet[][eq. (28)]{ver97a}.}, it increases the ensemble size from $m$ to $m + m_q$.  Consequently, to return to the
original ensemble size one needs to conduct ensemble size reduction at the end of the cycle.  If the ensemble size is
equal to or exceeds the dimension of the model subspace, such a reduction can be done losslessly; otherwise it is lossy.
The reduction of the ensemble size is conducted on line 21 of Algorithm~\ref{alg:ienkfq}.  Multiplication by $\sqrt{m -
  1}$ performs re-normalisation of $\mb A_2$ back to the standard EnKF form (\ref{A-enkf}).

A possible way of reducing the ensemble size to $m$ is to keep the $m - 1$ largest principal components of $\mb
A_2^\mathrm{a}$ and use the remaining degree of freedom to centre the reduced ensemble to zero.  In practice the
magnitude of ensemble members produced by this procedure can be quite non-uniform, similar to that of the SVD spectrum
of the ensemble.  This can have a detrimental effect on performance in a nonlinear system; therefore, one may need to
apply random mean-preserving rotations to the ensemble to render the ensemble distribution more Gaussian.

This reduction, based on the SVD, actually represents a \emph{marginalisation}, in a probabilistic sense, over all the
remaining degrees of freedom.  Assuming Gaussian statistics of the perturbations, the marginalisation can be rigorously
performed this way as the excluded modes are orthogonal to the posterior ensemble subspace.  In the limit of the
Gaussian approximation, this guarantees that the reduction to the posterior ensemble space accounts for all information
available in this ensemble space.

In the IEnKF-Q algorithm, the computational cost induced by $m_q$ is due to the cost of the observation operator to be
applied to the $m_q$ additional members of the ensemble in the analysis, as seen in \eqref{zerograd}, \eqref{A} and
\eqref{H}.  In contrast, the cost associated to the $m$ members of the ensemble is due to the application of both the
evolution model and observation operators, which is potentially much greater, as seen in \eqref{zerograd}, \eqref{A},
\eqref{H} and \eqref{M}.  A large $m_q$ also potentially increases the computational cost of the nonlinear optimisation
of cost function \eqref{J}, which is nonetheless expected to be often marginal compared to the computational cost of the
models.  For realistic applications, $m_q$ could be chosen to be reasonably small by pointing $\mb A^q_2$ to the most
uncertain, possibly known a priori, directions of the model, such as the forcings.  These are often called stochastic
perturbations of the physical tendencies, see \citet{bui99}, section~2.5 of \citet{wu08a} and section~5.c of
\citet{hou09a}.  Because they are randomly selected, these perturbations are actually meant to explore a number of
independent model error directions greater than $m_q$ but over several cycles of the DA scheme.

\section{Numerical tests}
\label{sec:tests}

This section describes a number of numerical tests of the IEnKF-Q with the Lorenz-96 model \citep{lor98a} to verify its
performance against the IEnKF and EnKF.

The model is based on $40$ coupled ordinary differential equations in a periodic domain:
\begin{align}
  \label{l40}
  & \dot{x}_i = (x_{i+1} - x_{i-2}) \, x_{i-1} - x_i + 8, \quad i = 1,\dots,40;\\
  & x_0 = x_{40}, \quad x_{-1} = x_{39}, \quad x_{41} = x_1.
\end{align}
Following \citet{lor98a}, this system is integrated with the fourth order Runge-Kutta scheme, using a fixed time step of
$\delta t = 0.05$, which is considered to be one model step.  The model noise is added after integrating equations
(\ref{l40}) for the time length of each DA cycle. An alternative would be to gradually add model noise at each model time
step, which would be consistent if the original model was based on a continuous stochastic differential equation.  Even
though less elegant, we chose the former approach because, in that case, the actual model error covariance is guaranteed
to match the assumed model error covariance.  However, an implication of such approach is that the properties of the
resulting stochastic model depend on the length of the cycle.  This applies to the true model state as well as to the
ensemble members of the EnKF- and IEnKF-based methods to be defined later.

In the following twin experiments, each variable of the model is independently observed once per DA cycle with Gaussian
observation error of variance 1: $\mb R = \mb I$.  The performance metric we use is the filtering analysis root mean
square error (RMSE) averaged over $10^5$ cycles after a spinup of $5000$ cycles.  For each run the optimal inflation is
chosen out of the following set: $\{1,1.02,1.05,1.1,1.15,1.2,1.25,1.3,1.4,1.5,1.75,2,2.5,3,4\}$.

In the following experiments, we choose $m_q=41$ for the IEnKF-Q so that $\mb A^q_2$ can span the whole range of $\mb Q$
whatever its actual form, which ensures that \eqref{A2q} is exactly satisfied. This should highlight the full potential of
the IEnKF-Q.

As justified in section~\ref{sec:algorithm}, random mean-preserving rotations of the ensemble anomalies are sometimes
applied to the IEnKF-Q, typically in the very weak model error regime.

The performance of the IEnKF-Q is compared to that of the EnKF using the ensemble transform Kalman filter scheme
\citep[ETKF,][]{bis01a} modified to accommodate the additive model error.  The additive model error needs to be
accounted for after the propagation step, so as to have $\mb P_2^\mathrm{f} = \mb M\mb P_1^\mathrm{a} \mb M \T + \mb Q$.
However, in an ensemble framework where the ensemble size $m$ is smaller than the size of the state space $n$,
generally, one cannot have ensemble of anomalies $\mb A_2^\mathrm{f}$ such that $\mb A_2^\mathrm{f}(\mb
A_2^\mathrm{f})\T=\mb P_2^\mathrm{f}$.  To accommodate model error into the EnKF or IEnKF frameworks, we use two
modifications of each of these schemes, referred to as stochastic and deterministic approaches.

The stochastic approach is
\begin{align}
  \label{eq:rand}
  \mb A_2^\mathrm{f} = \mb M \mb A_1^\mathrm{a} + \mb Q^{1/2}\Xi,
\end{align}
where $\Xi$ is an $n \times m$ matrix whose columns are independently sampled from $\mathcal N(\mb 0,\mb I)$.  With
these anomalies, one has $\mathrm{E}\left[\mb A_2^\mathrm{f}(\mb A_2^\mathrm{f})\T \right]=\mb M\mb P_1^\mathrm{a} \mb M \T + \mb Q$.  When
applying this approach to the EnKF, we refer to it as EnKF-Rand. Be wary that the EnKF-Rand is not the original
stochastic EnKF; its analysis step is deterministic.

The deterministic approach is to substitute the full covariance matrix $\mb Q$ with its projection $\widehat{\mb Q}$
onto the ensemble subspace: $\widehat{\mb Q} = \mb \Pi_{\mb A} \mb Q \mb \Pi_{\mb A}$, where $\mb \Pi_{\mb A} = \mb A
\mb A^\dagger$ is the projector onto the subspace generated by the anomalies (the columns of) $\mb A = \mb M \mb A_1^\mathrm{a}
$, and $\mb A^\dagger$ denotes the Moore-Penrose inverse of $\mb A$.  This yields the factorisation (which is
approximate if $m \le n$):
\begin{align}
  \label{hatQ}
  \mb P_2^\mathrm{f} & \approx \mb M \mb P_1^\mathrm{a} \mb M \T  + \widehat{\mb Q}
  = \mb A \mb A \T + \mb A \mb A^\dagger \mb Q  (\mb A^\dagger)\T \mb A\T \nonumber \\
  & \approx \mb A \left[ \mb I +  \mb A^\dagger \mb Q  (\mb A^\dagger)\T \right] \mb A\T .
\end{align}
Hence, the anomalies that satisfy \eqref{hatQ} are \citep{raa15b}
\begin{align}
  \label{eq:det}
  \mb A_2^\mathrm{f} = \mb A \left[\mb I + \mb A^\dagger \mb Q (\mb A^\dagger)\T \right]^{1/2}.
\end{align}
When applying this approach to the EnKF, we refer to it as EnKF-Det.

Those two simple ways to add model noise to the analysis of the EnKF can also be applied to the standard IEnKF.
At each iteration, the IEnKF smoothing analysis at $t_1$, yielding $\mb A^\mathrm{s}_1$, is followed by either \eqref{eq:rand} or
\eqref{eq:det} with $\mb A^\mathrm{a}_1$ replaced with $\mb A^\mathrm{s}_1$, which yields the IEnKF-Rand and IEnKF-Det methods,
respectively.

These heuristic methods are fostered by the decoupling analysis in section~\ref{sec:decoupling} when $\mathcal H$ is
linear. This analysis suggests that the IEnKF smoothing analysis at $t_1$ should actually be performed with an
observation error covariance matrix $\mb R + \mb H \mb Q \mb H\T$ in place of $\mb R$.  Note, however, that we did not
observe any significant differences in the performance of the IEnKF-Rand and IEnKF-Det with or without this correction.

Further, it can be shown that the heuristic IEnKF-Rand yields
\begin{align}
  \label{eq:P2E}
  \mb P^\mathrm{a}_2   \approx & \mathrm{E} \left[ \mb M  \mb P_1^\mathrm{s} \mb M\T + \mb Q \right] \nonumber \\
 \approx &
  \mb Q + \mb M \mb P_1^\mathrm{a} \mb M\T - \mb M \mb P_1^\mathrm{a} \mb M\T \mb H\T \nonumber \\
  &\times \left[ \mb R + \mb H (\mb Q + \mb M \mb P_1^\mathrm{a} \mb M\T)\mb H\T
    \right]^{-1} \mb H \mb M \mb P_1^\mathrm{a} \mb M\T ,
\end{align}
whereas the rigorous IEnKF-Q yields
\begin{align}
  \label{eq:P2}
  & \mb P^\mathrm{a}_2  =  \mb Q + \mb M \mb P_1^\mathrm{a} \mb M\T - (\mb Q + \mb M \mb P_1^\mathrm{a} \mb M\T)\mb H\T  \nonumber \\
  & \times \left[ \mb R + \mb H (\mb Q + \mb M \mb P_1^\mathrm{a} \mb M\T)\mb H\T \right]^{-1} \mb H (\mb Q + \mb M \mb P_1^\mathrm{a} \mb M\T) .
\end{align}
These posterior error covariance matrices \eqref{eq:P2E} and \eqref{eq:P2} are very similar (although objectively
different whenever $\mb Q \neq \mb 0$), so that the IEnKF-Rand and IEnKF-Det can be considered as relevant
approximations of the IEnKF-Q.  Note that if $m \ge n+1$, \eqref{eq:P2E} provides an exact expression for $\mb P^\mathrm{a}_2$ of
the IEnKF-Det.  The tests below show that with a full-rank (or nearly full-rank) ensemble and tuned inflation the
IEnKF-Det and IEnKF-Q can yield very similar performance.

Finally, let us mention that the IEnKF could also use one of the advanced model error perturbation schemes introduced by
\citet{raa15b}, with the goal to form other IEnKF-based approximate schemes of the IEnKF-Q.  Yet, we do not expect these
alternative schemes to fundamentally change the conclusions to be drawn from the rest of this study.

\subsection{Test 1: nonlinearity}
\label{sec:test1}

This test investigates the performance of the schemes depending on the time interval between observations, covering DA
regimes from weakly nonlinear to significantly nonlinear.  The ensemble size is $m = 20$, chosen so that it is greater
than the dimension of the unstable-neutral subspace (which is here $14$ and to which we add $1$ to account for the
redundancy in the anomalies) and hence avoids the need for localisation. Each model variable is observed once at each DA
cycle.  Model error covariance is set to $\mb Q = 0.01 \, T \, \mb I $, where $T$ is the time interval between
observations in units of the model time-step $\delta t=0.05$.  For instance, $T=4$ stands for $4 \times \delta t = 0.20$
units of the Lorenz-96 model.  It is therefore proportional to the cycle length.  Since it is added after model
integration over the cycle length, the model error variance per unit of time is kept constant.  Even though this value
of $\mb Q$ is two orders of magnitude smaller than $\mb R$ when $T=1$, we found it to be realistic.  Indeed, the
standard deviation of the perturbation added to the truth of the synthetic experiment is $0.1$ in that case, to be
compared to a root mean square error of about $0.2$ obtained for the analysis of the EnKF in a perfect model experiment
with $T=1$, the Lorenz-96 model, and the data assimilation setup subsequently described.

\begin{figure}[!ht]
  \centering
  \includegraphics[clip=true,width=\columnwidth]{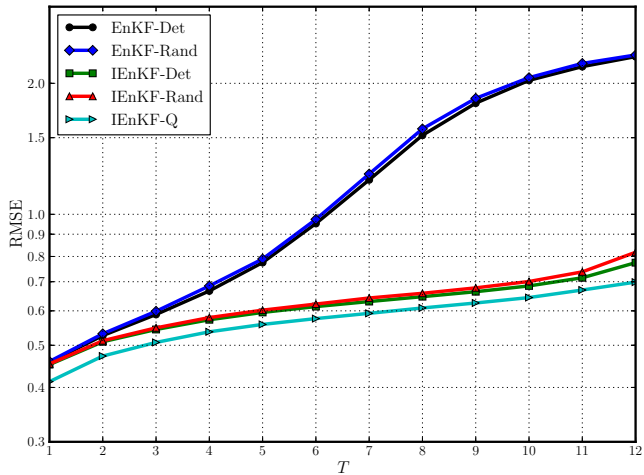}
  \caption{Test 1: dependence of the mean analysis RMSE on the time interval between observations $T$ in units of $\delta t$.  $m = 20$, $\mb Q = 0.01 T \mb I$.}
  \label{test1}
\end{figure}

Figure~\ref{test1} compares the performance of the EnKF, the IEnKF and the IEnKF-Q depending on the time interval $T$
(in units of $\delta t$) between observations.  From $T = 3$, the iterative methods noticeably outperform the EnKF due
to the increasing nonlinearity.  For all $T$, the IEnKF-Q consistently outperforms the IEnKF-Rand and IEnKF-Det.
Interestingly, this conclusion holds for the weakly nonlinear case $T = 1$.  The reason is that the IEnKF-Q internally
uses ensemble of size $m + m_q$ during the minimisation, while the other methods use only ensembles of size $m$.  As
will be shown in sections~\ref{sec:test2} and \ref{sec:test3} (Tests 2 and 3), this advantage decreases with larger
ensembles.

\begin{figure}[!ht]
  \centering
  \includegraphics[clip=true,width=\columnwidth]{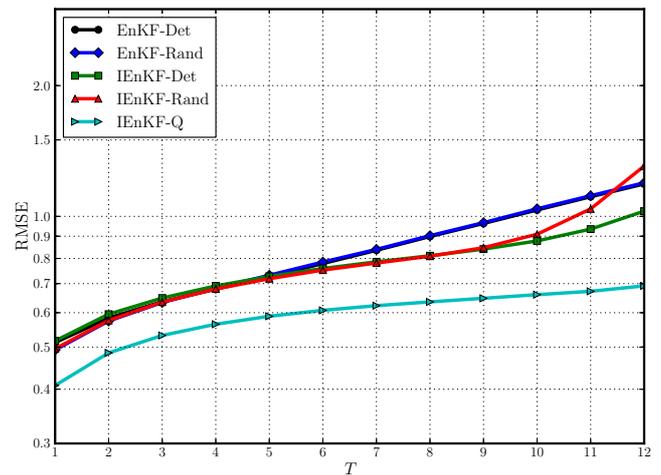}
  \caption{Test 1: dependence of the mean analysis RMSE on the time interval between observations $T$ (in units of $\delta t$) with settings
    similar to those from \citet{raa15b}, their Figure~4; $m = 30$, and $\mb Q$ is non-diagonal (see text).}
  \label{test1-raanes}
\end{figure}

Figure~\ref{test1-raanes} replicates the settings used for Figure~4 of \citet{raa15b}.  Our EnKF-Rand and EnKF-Det
schemes correspond to their Add-Q and Sqrt-Core, respectively.  Note that \citet{raa15b} added model error with
covariance $[\mb Q]_{ij} = 0.05 \left( \exp[-d^2(i,j)/30] + 0.1\delta_{ij}\right)$ after each model step, where
$\delta_{ij}$ is the Kronecker symbol and $d$ is the distance on the circle: $d(i,j) = \min( |i-j|, 40- |i-j|)$.
Compared to Figure~\ref{test1}, the ensemble size is, accordingly, increased to $30$, and the model error covariance is
increased and correlated.  It can be seen that the increase in the model error results in a more pronounced advantage of
the IEnKF-Q over the other methods.  The relative performance of the non-iterative schemes is better than in
Figure~\ref{test1} because of the increased ensemble size.

\subsection{Test 2: model noise magnitude}
\label{sec:test2}

This test investigates the relative performance of the schemes depending on the magnitude of model error both in a
weakly nonlinear and significantly nonlinear case.  The tests for all schemes involved are conducted with ensemble size $m =
20$. Moreover, results with full-rank ensemble $m = 41$ are shown for IEnKF-Det and IEnKF-Q.

\begin{figure}[!ht]
  \centering
  \includegraphics[clip=true,width=\columnwidth]{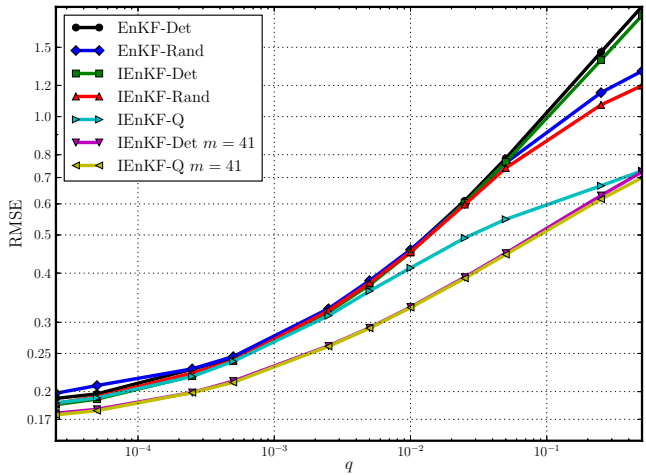}
  \caption{Test 2: dependence of the mean analysis RMSE on the magnitude of model error in a weakly nonlinear case $T = 1$; $\mb Q = q T \mb I$, $m = 20$.}
  \label{test2-1}
\end{figure}

Figure~\ref{test2-1} shows the performance of the schemes in a weakly nonlinear case $T = 1$.  For small model error $q
\lesssim 3\cdot 10^{-3}$, all schemes perform similarly; for $q \gtrsim 3\cdot 10^{-3}$ the IEnKF-Q starts to outperform
other schemes; and from $q \gtrsim 5 \cdot 10^{-2}$ the iterative schemes IEnKF-Rand and IEnKF-Det start to outperform
their non-iterative counterparts EnKF-Rand and EnKF-Det.

Interestingly, the IEnKF-Q does not show advantage over IEnKF-Det when both use full-rank ensembles $m=41$ (except,
perhaps, some very marginal advantage for larger model error $q \gtrsim 0.1$).  At a heuristic level, both schemes
indeed explore the same model error directions.  At a mathematical level, there are indications in favour of this
behaviour, including the marginal advantage.  Thanks to the decoupling analysis, we see that when $m=41$ the smoothing
analysis at $t_1$ of the IEnKF-Q and that of the IEnKF-Det become equivalent.  However, the filtering analysis at $t_2$
of the IEnKF-Q, \eqref{eq:MAP2}, is different albeit close to that of the IEnKF-Det, which is just the forecast of $\mb
x_1^\star$, i.e., the first term of \eqref{eq:MAP2}.  Concerning the update of the anomalies of the ensemble, we have
seen that \eqref{eq:P2E} is the $\mb P^\mathrm{a}_2$ of the IEnKF-Det when $m=41$, and is very close to the expression
of $\mb P^\mathrm{a}_2$ for the IEnKF-Q, \eqref{eq:P2}, except maybe when $\mb Q$ is large.

\begin{figure}[!ht]
  \centering
  \includegraphics[clip=true,width=\columnwidth]{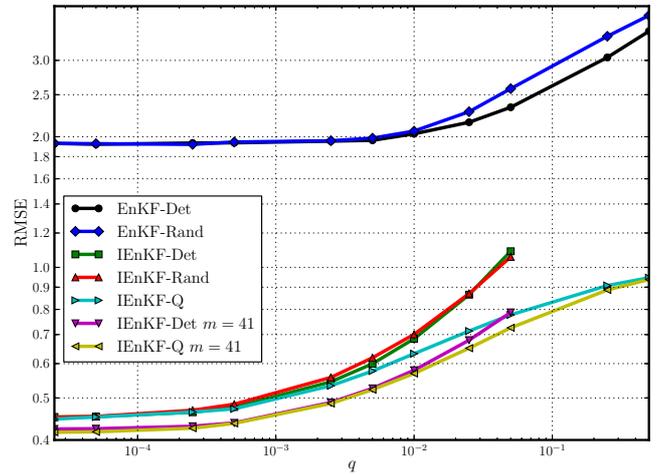}
  \caption{Test 2: dependence of the mean analysis RMSE on the magnitude of model error in a significantly nonlinear case $T = 10$; $\mb Q = q T \mb I$, $m = 20$.}
  \label{test2-10}
\end{figure}

In the significantly nonlinear case ($T=10$, Figure~\ref{test2-10}), the non-iterative schemes are no longer able to
constrain the model.  The empirically modified iterative schemes IEnKF-Rand and IEnKF-Det yield performance similar to
the IEnKF-Q up to $q \lesssim 2 \cdot 10^{-3}$ with the ensemble size $m = 20$, and up to $q \lesssim 10^{-2}$ with the
ensemble size $m = 41$; however, apart from underperforming the IEnKF-Q for larger model errors, they also lose
stability and are unable to complete $10^5$ cycles necessary for completion of these runs.  Interestingly, for very
large model error $q = 0.5$ ($\mb Q = 5\mb I$) the IEnKF-Q yields similar performance with the ensemble size of $m = 20$
and $m = 41$, with the RMSE $\sim 0.94$ much smaller than the average magnitude of the model error $\sim 2.2$. In this
regime, the analysis RMSE remains smaller than that obtained with the sole observations (estimated to be $0.994 \lesssim
1$ by A. Farchi, personal communication). This could be due to the variational analysis which spans the full state space
since $m_q = 41$, and, in this regime, little depends on the prior perturbations $\mb A^\mathrm{a}_1$.

\subsection{Test 3: ensemble size}
\label{sec:test3}

\begin{figure}[!ht]
  \centering
  \includegraphics[clip=true,width=\columnwidth]{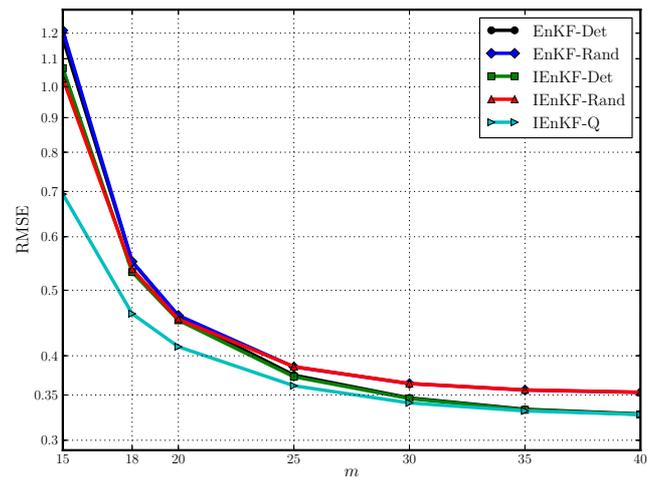}
  \caption{Test 3: dependence of the mean analysis RMSE on the ensemble size. $T = 1$, $\mb Q = 0.01 T \mb I$.}
  \label{test3-1}
\end{figure}

\begin{figure}
  \centering
  \includegraphics[clip=true,width=\columnwidth]{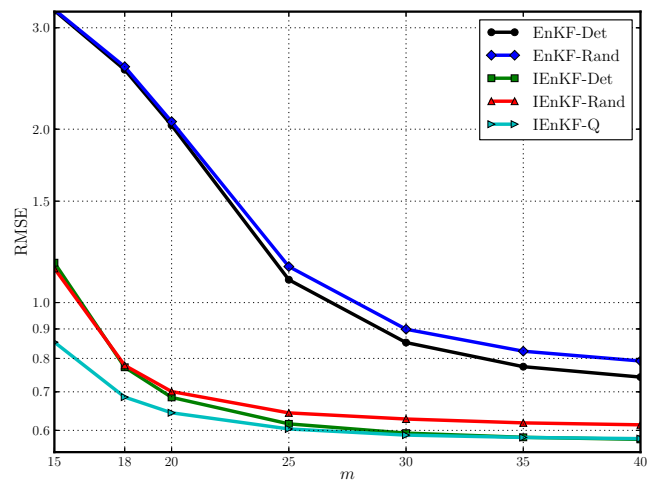}
  \caption{Test 3: dependence of the mean analysis RMSE on the ensemble size. $T = 10$, $\mb Q = 0.01 T \mb I$.}
  \label{test3-10}
\end{figure}

This test investigates the performance of the methods depending on the ensemble size both in a weakly nonlinear ($T =
1$, Figure~\ref{test3-1}) and a significantly nonlinear ($T = 10$, Figure~\ref{test3-10}) case.  The model error is set to a
moderate magnitude $\mb Q = 0.01 T \mb I$.

In line with the results of Test 2, we observe that the non-iterative schemes do not perform well in the significantly
nonlinear case.  The IEnKF-Q outperforms the other schemes when using smaller ensembles, but yields a performance
similar to that of the IEnKF-Det with a full-rank (or nearly full-rank) ensemble.  This is mainly due to its
search for the optimal analysis state over a large subspace. Likewise, the performance of the IEnKF-Q degrades when
restricting the model error directions to that of the ensemble space (yielding $m_q=m$), and yet remains slightly better
than the IEnKF-Det (not shown).

\subsection{Test 4: localisation}

A couple of numerical experiments are carried out to check that the IEnKF-Q can be made local. However, a detailed
discussion of the results is out of scope, since our primary concern is only to confirm the feasibility of a local
IEnKF-Q.  To this end, we have merged the local analysis as described in \citet{boc16a} and initially meant for the
IEnKF in perfect model conditions, with the IEnKF-Q algorithm. The resulting local IEnKF-Q algorithm is given in
Appendix A.

\begin{figure}[!ht]
  \centering
  \includegraphics[clip=true,width=\columnwidth]{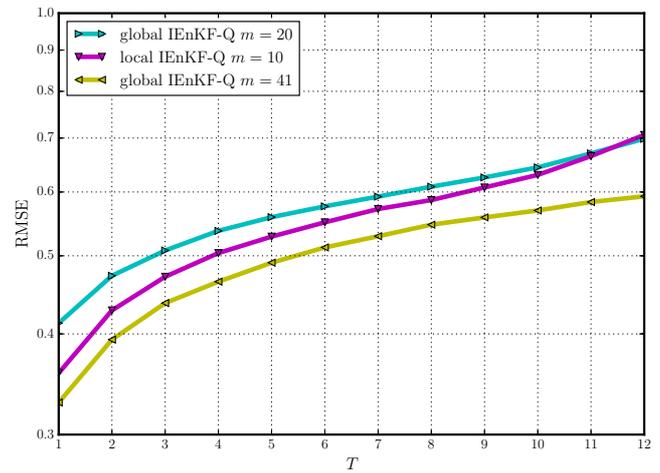}
  \caption{Test 4: dependence of the mean analysis RMSE on the time interval between observations $T$ in units of $\delta t$; $\mb Q = 0.01 T \mb I$.}
  \label{test4-1}
\end{figure}

First, we use the same experimental setup as for Figure~\ref{test1}, i.e. the RMSE as a function of $T$. In addition, we
consider the local IEnKF-Q with an ensemble size of $m=10$, which requires the use of localisation. The localisation
length is $10$ grid points (see Appendix A for its definition).  Dynamically covariant localisation is used \citep[see
  section 4.3 in][]{boc16a}. The RMSEs are plotted in Figure~\ref{test4-1}. From Figure~\ref{test1}, we transfer the
RMSE curve of the global IEnKF-Q with ensemble size $m=20$. The same RMSE curve but with $m=41$ was computed and added
to the plot.  The local IEnKF-Q RMSE curve lies in between those for the global IEnKF-Q with $m=20$ and $m=41$.

\begin{figure}[!ht]
  \centering
  \includegraphics[clip=true,width=\columnwidth]{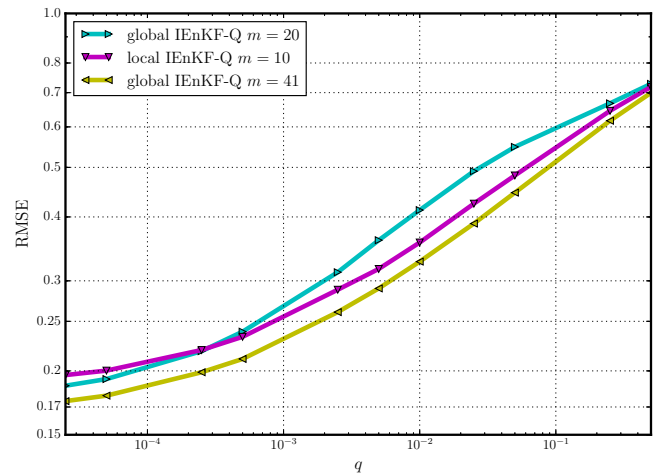}
  \caption{Test 4: dependence of the mean analysis RMSE on the magnitude of model error in a weakly nonlinear case $T = 1$; $\mb Q = q T \mb I$.}
  \label{test4-2}
\end{figure}

Second, we use the same experimental setup as for Figure~\ref{test2-1}, i.e. the RMSE as a function of the model error
magnitude.  From Figure~\ref{test2-1}, we transfer the RMSE curves of the global IEnKF-Q with ensemble size $m=20$ and
$m=41$.  In addition, we consider the local IEnKF-Q with an ensemble size of $m=10$, which requires the use of
localisation. Again, the localisation length is $10$ grid points. The RMSEs are plotted in Figure~\ref{test4-2}.  The
local IEnKF-Q RMSE curve lies in between those for the global IEnKF-Q with $m=20$ and $m=41$, with a slight
deterioration for very weak model error.

Both tests show that a local IEnKF-Q is not only feasible but also yields very accurate results, with a local
$10$-member implementation outperforming a global $20$-member implementation.

\section{Discussion}
\label{sec:discussion}

The presence of model error in a DA system causes lossy transmission of information in time.  The remote in time
observations have less impact on the model state estimates compared to the perfect-model case; and conversely, the
current observations have relatively more impact.  The latter follows from the KF solution, $\mb P^\mathrm{f}_{i+1} =
\mb M_{i+1} \mb P^\mathrm{a}_i \mb M_{i+1}\T + \mb Q_{i+1}$, which increases the forecast covariance by the model error
covariance; therefore the presence of model error shifts the balance between the model state and observations in the
analysis towards observations.  The former can be seen from the ``decoupled'' equation (\ref{u-lin}), when the smoothed
state $\mb x_1^\mathrm{s}$ can be obtained essentially in the perfect-model framework using increased observation error
$\mb R$ according to (\ref{Ru}).

This dampened transmission of information shuts down the usual mechanisms of communication in perfect-model linear EnKF
systems, such as applying calculated ensemble transforms at a different time or concatenating ensemble observation
anomalies within observation window.  Because the IEnKF is based on using observations at time $t_2$ for updating the
system state at time $t_1$, it was not intuitively clear whether it could be rigorously extended for the case of
imperfect model.  Fortunately, the answer to this question has proved to be positive.  Moreover, the form of the IEnKF-Q
solution (\ref{w-sol}) suggests that it may be possible to further generalise its framework to assimilate asynchronous
observations (i.e, observations collected at different times) within a DA cycle.

In practice, the concept of additive model error is rarely directly applicable, for two reasons.  Firstly, the often
encountered model errors such as random or some systematic forcing errors, representativeness errors, errors in
parametrisations and basic equations and so on are non-additive by nature.  Secondly, even if the model error is
additive, it is generally difficult to characterise.  Because in the KF the model error covariance is
directly added to the forecast error covariance, a misspecification of model error will result in a suboptimal, and
possibly unbalanced, analysis.

Nevertheless, the additive model error is an important theoretical concept because it permits exact linear recursive
solutions known as Kalman filter and Kalman smoother as well as treatment by means of control theory (4D-Var).
Furthermore, in 4D-Var the additive model error can be used empirically for regularisation of the minimisation problem
that becomes unstable for long assimilation windows \citep[e.g.,][p. 451]{bla14a}.  Therefore, there may be potential
for empirical use of the additive model error in the EnKF to improve numerics.  It indeed can often be perfectly
feasible to specify some sort of additive model error as a tuneable parameter of a suboptimal system, similarly to the
common use of inflation.  In fact, a number of studies found that using empirical additive model error in EnKF systems,
alone or in combination with inflation, can yield better performance than using inflation only \citep[e.g.,][]{whi08a}.
Another example of employing model error in a suboptimal system is using (so far without marked success) the hybrid
covariance factorised by ensemble anomalies augmenting a small (rank deficient) dynamic ensemble and a large static
ensemble \citep{cou09a}.

In this study, we have assumed that model error statistics are known.  In some simple situations, these could be
estimated online with techniques such as those developed by \citet{tod15a}.  Nonetheless, using these empirical Bayesian
estimation techniques here would have obscured the methodological introduction to the IEnKF-Q.

\section{Summary}
\label{sec:summary}

This study proposes a new method called IEnKF-Q that extends the iterative ensemble Kalman filter (IEnKF) to the case of
additive model error.  The method consists of a Gauss-Newton minimisation of the nonlinear cost function conducted in
ensemble space spanned by the propagated ensemble anomalies and anomalies of the model error ensemble.  To simplify the
algebraic form of the Gauss-Newton minimisation, the IEnKF-Q concatenates the expansion coefficients $\mb u$ and $\mb v$
into a single vector $\mb w$, and augments ensemble anomalies $\mb M \mb A_1^\mathrm{a}$ and $\mb A_2^q$ into a single
ensemble $\mb A$.  After that, the minimisation takes the form (\ref{w-sol}) similar to that in the perfect-model case.

Algorithmically, the method can take many variations including ``transform'' and ``bundle'' versions, and various
localisation approaches.  An example algorithm suitable for low dimensional systems is presented in
section~\ref{sec:algorithm}.  Using this algorithm, the method is tested in section~\ref{sec:tests} in a number of
experiments with the Lorenz-96 model.  In all experiments, the IEnKF-Q outperforms both the EnKF and IEnKF adapted for
handling model error either in a ``stochastic'' or ``deterministic'' way, except in situations with full-rank ensemble
and weak to moderate model error, where it performs equally with the IEnKF-Det.  Surprisingly, it also outperforms these
methods in weakly nonlinear situations, when the solution is essentially found at the very first iteration, and
iterative schemes should not have any marked advantage over non-iterative schemes.  This is caused by using full-rank
(augmented) forecast ensemble anomalies in the analysis, and only reducing the ensemble size back to the initial one at
the very end of the cycle.  Note that in practice the cost of the IEnKF-Q in high-dimensional systems can be expected to
be similar to that of the IEnKF because both methods use ensembles of the same size in propagation.

One interesting feature of the IEnKF-Q is the decoupling of iterations over $\mb u$ and $\mb v$ made possible in
presence of a linear observation operator.  In this case, $\mb u$ can be found using the (perfect-model) IEnKF with an
increased observation error covariance, followed by obtaining $\mb v$ in a single iteration.  The decoupling can be the
underlying reason why in certain situations the IEnKF-Q and IEnKF-Det show equal performance.

\acks
The authors would like to thank two anonymous reviewers for their useful comments and suggestions.  They are grateful to
T. Janji\'c for her invitation to submit this work.  M. Bocquet and J.-M. Haussaire acknowledge the contribution of
INSU via the LEFE/MANU DAVE project.  CEREA is a member of the Institut Pierre Simon Laplace (IPSL).

\section*{Appendix A: Local IEnKF-Q algorithm}
\label{sec:appendixA}

Algorithm \ref{alg:lienkfq} corresponds to a local analysis variant of the global IEnKF-Q.  It stems from merging
Algorithm \ref{alg:ienkfq} with the local scheme described in Table 2 of \citet{boc16a}.  The local analyses are looped
over the space grid points $i=1,\ldots,n$.  Each local analysis uses a local observation error covariance matrix $\mb
R^i$ whose inverse has been tapered with the Gaspari-Cohn piecewise rational function \citep[equation (4.10)
  in][]{gas99}, where the localisation length is defined to be their $c$ parameter.

\begin{algorithm}
%%  \setstretch{1.35}
  \caption{\label{alg:lienkfq} A local analysis and transform variant of the IEnKF-Q, using a Gauss-Newton minimisation.
``$\mathrm{SR}(\mb A, m)$'' denotes ensemble size reduction from  $m+m_q$ to $m$.}
  \begin{algorithmic}[1]
\Function{$[\mb E_2]$ = \mbox{\bf lienkf\_cycle}}{$\mb E^\mathrm{a}_1, \, \{\mb A_i^q\}_{i=1,\ldots,n}, \, \mb y_2, \newline \, \{\mb R^i\}_{i=1,\ldots,n}, \, \mathcal M, \mathcal H$}
    \State $\mb x_1^\mathrm{a} = \mb E_1^\mathrm{a} \, \mb 1 / m$
    \State $\mb A_1^\mathrm{a} = (\mb E_1^\mathrm{a} - \, \mb x_1^\mathrm{a} \mb 1^{\mathrm T}) / \sqrt{m - 1}$
    \For{$i=1,\ldots, n$}
    \State $\mb D^i = \mb I, \quad \mb w^i = \mb 0$
    \EndFor
    \Repeat
    \For{$i=1,\ldots, n$}
    \State $[\mb x_1]_i = [\mb x_1^\mathrm{a}]_i + [\mb A_1^\mathrm{a} \mb w^i_{1:m}]_i$
    \State $\mb T^i = ([\mb D^i]_{1:m,1:m})^{1/2}$
    \State $[\mb E_1]_i = [\mb x_1 \mb 1^{\mathrm T}]_i + [\mb A_1^\mathrm{a} \mb T^i]_i \sqrt{m - 1}$
    \EndFor
    \State $\mb E_2 = \mathcal M (\mb E_1)$
    \For{$i=1,\ldots, n$}
    \State $[\mb x_2]_i = [\mb E_2 \mb 1/m]_i + [\mb A^q_i \mb w^i_{m+1:m+m_q}]_i$
    \EndFor
    \For{$i=1,\ldots, n$}    
    \State $\mb H \mb A_2^i = \mathcal H(\mb E_2)(\mb I - \mb 1\mb 1^{\mathrm T}/ \, m) \, (\mb T^i)^{-1}/ \sqrt{m - 1}$
    \State ${\mb H \mb A^q_i = \mathcal H(\mb E_2 \mb 1 \mb 1^{\mathrm T}/ \, m + \mb A^q_i \sqrt{m_q - 1})} $
    \Statex \hspace{3.5cm}
           $ \times (\mb I - \mb 1 \mb 1^{\mathrm T} / \, m_q) / \sqrt{m_q - 1}$
    \State $\mb H \mb A^i = [\mb H \mb A^i_2, \mb H \mb A_i^q]$
    \State $\nabla \! \mathcal J^i =  \mb w^i - (\mb H\mb A^i)^{\mathrm T} (\mb R^i)^{-1} [\mb y_2 - \mathcal H(\mb x_2)]$
    \State $\mb D^i = [\mb I + (\mb H\mb A^i)^{\mathrm T} (\mb R^i)^{-1} \mb H\mb A^i]^{-1}$
    \State $\Delta \mb w^i = -\mb D^i \, \nabla \! \mathcal J^i$
    \State $\mb w^i := \mb w^i + \Delta \mb w^i $
    \EndFor
    \Until $\sum\limits_{i=1}^{n}\| \Delta \mb w^i \| < n\varepsilon$
    \For{$i=1,\ldots, n$}    
    \State $\mb A_2^i = \mb E_2 \, (\mb I - \mb 1 \mb 1^{\mathrm T} / \, m) \, (\mb T^i)^{-1}$
    \State $\mb A^i = [\mb A_2^i / \sqrt{m - 1}, \mb A_i^q] \, (\mb D^i)^{1/2}$
    \State $\mb A^i_2 = \mathrm{SR}(\mb A^i, m) \sqrt{m  - 1}$
    \State $[\mb E_2]_i = [\mb x_2 \mb 1^{\mathrm T}]_i + [\mb A^i_2]_i$
    \EndFor
    \EndFunction
  \end{algorithmic}
\end{algorithm}

\bibliographystyle{wileyqj}
\bibliography{refs}

\end{document}